\documentclass[aps,prr,superscriptaddress,twocolumn]{revtex4-2}
\usepackage{bm}
\usepackage{amsmath}
\usepackage{amsfonts}
\usepackage{amssymb}
\usepackage{graphicx}
\usepackage{color}
\usepackage[colorlinks,linkcolor=blue,anchorcolor=blue,citecolor=blue,urlcolor=blue]{hyperref}
\usepackage{dcolumn}

\begin{document}

\title{Supercurrent-carrying supersolid in spin-orbit-coupled Bose-Einstein condensates}

\author{Hao~Lyu}
\email{lyuhao@shu.edu.cn}
\affiliation{Institute for Quantum Science and Technology, Department of Physics, Shanghai University, Shanghai 200444, China}
\affiliation{Quantum Systems Unit, Okinawa Institute of Science and Technology Graduate University, Onna, Okinawa 904-0495, Japan}

\author{Yuanyuan~Chen}
\affiliation{Institute for Quantum Science and Technology, Department of Physics, Shanghai University, Shanghai 200444, China}

\author{Qizhong Zhu}
\affiliation{Guangdong Basic Research Center of Excellence for Structure and Fundamental Interactions of Matter, Guangdong Provincial Key Laboratory of Quantum Engineering and Quantum Materials, School of Physics, South China Normal University, Guangzhou 510006, China}
\affiliation{Guangdong-Hong Kong Joint Laboratory of Quantum Matter, Frontier
	Research Institute for Physics, South China Normal University, Guangzhou
	510006, China}

\author{Yongping~Zhang}
\email{yongping11@t.shu.edu.cn}
\affiliation{Institute for Quantum Science and Technology, Department of Physics, Shanghai University, Shanghai 200444, China}

\begin{abstract}
	
One of brilliant achievements in spin-orbit-coupled Bose-Einstein condensates is the discovery and observation of the  supersolid stripe states. So far, all studied supersolid stripe states do not carry supercurrent.  In this paper, we reveal the existence of supercurrent-carrying supersolids in spin-orbit-coupled Bose-Einstein condensates.  
The supersolid family has a parabolic-like dispersion and carries supercurrent which is proportional to the quasimomentum.  Energetic and dynamical instabilities can break supercurrent-carrying ability of this supersolid family.  An insightful interpretation of the dynamical instability of supercurrent-carrying supersolids from the pure plane-wave phase is provided.

\end{abstract}

\maketitle

\section{Introduction}

In the recent years, synthetic spin-orbit coupling in ultracold atoms has been recognized as a powerful tool to simulate many-body physics and explore exotic superfluids~\cite{Goldman,Zhai2015,Zhang}. The spin-orbit coupling can be artificially introduced in neutral atoms by coupling two hyperfine states via a pair of Raman lasers, 
where the two states are treated as pseudospin~\cite{Lin,Cheuk,Wang}.
The celebrated spin-orbit coupled Bose-Einstein condensates (BECs) possess unconventional and rich ground-state phase diagrams
~\cite{Zhu,Martone,LiY,Zheng,Hamner2014,Hamner,WuZ,Campbell,LiCH,Valdes-Curiel}.  
Outstanding exotic ground states of such a system include the plane-wave phase and the stripe phase. 
The plane-wave phase breaks the time-reversal symmetry so that it is spin-polarized~\cite{Wu2011,Martone}. 
The fundamental feature is that the plane-wave phase supports a phonon-maxon-roton structure in its elementary excitation spectrum~\cite{Ji, Khamehchi2014}. This phase provides a unique route to generate rotons~\cite{Ozawa,Khamehchi,Yi}.  The stripe phase simultaneously breaks the continuous translational symmetry and the gauge symmetry and accordingly possesses supersolidity~\cite{Ho,Wang2010,LiY2013,ChenX,LiJR}.  
In the stripe phase, the density crystalline structure offers an intriguing approach for experimental observation,
while the elementary excitation spectrum has Bloch band-gap structures, which includes two gapless Nambu-Goldstone modes in the long wave-length regime. The theoretical discovery~\cite{Wang2010,Ho,LiY2013} and experimental observation~\cite{LiJR} of the supersolid stripe phase constitute a main achievement in the field of spin-orbit-coupled BECs.
Rich physics relevant to supersolid stripes have been  revealed, including dynamics and excitations in these exotic states~\cite{Bersano,Hu2012, Hou,Lyu2020,LiGQ,Geier}.
On the other hand, atomic BECs with long-range interactions are active platforms to support roton excitations and to generate supersolid phases~\cite{Tanzi,Gallemi,Chomaz2023}.  Superfluid properties of long-range-interaction-induced supersolids have attracted a great deal of research attention~\cite{Macri,Cinti,Wessel,Saccani}. The spin-orbit-coupling-induced supersolid stripes provide an alternative means to detect the exotic superfluidity.

The fundamental property of superfluids is that they can carry current flow without dissipation. The supercurrent-carrying ability is limited by two different instabilities~\cite{Pitaevskii}.  Energetic instability happens when elementary excitations of a supercurrent-carrying superfluid have negative energies. The instability physically relates to the Laudau's criterion of superfluidity. When the velocity of the superfluid is larger than a critical velocity, the superfluid becomes energetically unfavourable and then loses the supercurrent-carrying ability.  Dynamical instability emerges when imaginary energies appear in elementary excitations~\cite{Wu}. The instability causes the exponential growth of excitation perturbations and therefore breaks superfluidity.   
Immediately after the first experimental realization of spin-orbit-coupled BECs~\cite{Lin}, the existence of supercurrent-carrying plane-wave states are proposed~\cite{Zhu,Ozawa,Zhang2019}. 
Dynamical instability of these states is caused by their negative effective mass, 
and the energetic instability is shown to relate to the negative-energy phonon or negative-energy roton~\cite{Ozawa}.  
So far, all studied supersolid stripe phases in spin-orbit-coupled BECs in literature do not carry supercurrent. Whether there exist supercurrent-carrying supersolid stripes becomes an intriguing question. The main intuitive concern is that the density crystalline structure of stripes is not in favour of current flow. If they exist, the instabilities of their superfluidity naturally arise as a further question.

In this paper, we address the above two questions by constructing supercurrent-carrying supersolid stripes and analyzing their stabilities.  
The construction is inspired by Bloch waves in optical lattices~\cite{Denschlag,Eiermann,Martone2017,Geiger}.  
In a one-dimensional optical lattice, the Bloch wave function is $\exp(ikx)\phi(x)$, 
where $k$ is the quasimomentum, and $\phi(x)$ is a periodic function. It is well known that the Bloch waves carry current flow if $k\ne 0$. Therefore, the plane-wave prefactor $\exp(ikx)$ becomes the source of the current carried by the Bloch waves.  We assume the wave function of supersolids to have a similar form as Bloch waves. The obvious difference is that the periods of the supersolid wave functions should be determined by minimizing the associated energy functional, 
while the period of the Bloch waves in optical lattice has a fixed value related to the wave length of the lattice beams.  
The supersolid family we found has the same density period and carries current flow that is proportional to the quasimomentum.  
The features of the supersolid family, including contrasts and spin polarizations, are identified.  
Energetic and dynamical instabilities are revealed, which can lead to the breakdown of their supercurrent-carrying ability. We calculate the relevant elementary excitations of the supersolid family, from which the energetic and dynamical instabilities are identified. The energetic instability relates to the well-known Landau's criterion of superfluidity. The dynamical instability is qualitatively understood from the pure plane-wave phase.  In spin-orbit-coupled BEC experiments, the detuning is a tunable and important parameter. We also study the supercurrent-carrying supersolid stripes in the presence of the detuning.

The outline of the paper is as follows.
In Sec.~\ref{model}, we present our theoretical frame for investigating supercurrent-carrying supersolids and their instabilities. 
In Sec.~\ref{nodetuning}, we reveal the existence of the supercurrent-carrying supersolid family in the absence of detuning.  
Important features of the supersolid family are identified.  The reasons of their energetic and dynamical instabilities are explained.  The supercurrent-carrying supersolid stripes in the presence of detuning are considered in Sec.~\ref{detuning}. 
In Sec.~\ref{exp}, we discuss experimental accessibility of the novel supercurrent-carrying supersolids.
Finally, the conclusions follow in Sec.~\ref{conclusion}.

\section{Theoretical model}
\label{model}

We consider a spin-orbit-coupled spin-1/2 BEC.  
The spin-orbit coupling can be induced between two hyperfine states of the atoms via a two-photon transition process induced by a pair of Raman beams~\cite{Lin}. The system is described by the Gross-Pitaevskii (GP) equations,
\begin{align}
i\frac{\partial\psi}{\partial t}=\left(H_{\mathrm{SOC}}+H_\mathrm{int} \right)\psi,
\label{eq:motion}
\end{align}
where $\psi=(\psi_1, \psi_2)^T$ is the two-component wave function. The GP equations are dimensionless. We choose the units of momentum, length, and energy as $\hbar k_{\mathrm{Ram}}$, $1/k_{\mathrm{Ram}}$, and $\hbar^2 k^2_{\mathrm{Ram}}/m$, respectively. 
Here, $k_{\mathrm{Ram}}=2\pi/\lambda_{\mathrm{Ram}}$ is the wave vector of the Raman lasers with $\lambda_{\mathrm{Ram}}$ being the corresponding wave length, and $m$ is the atom mass. The spin-orbit-coupled single-particle Hamiltonian is
\begin{equation}
H_{\mathrm{SOC}}=-\frac{1}{2}\frac{\partial^2}{\partial x^2}-i\frac{\partial}{\partial x}\sigma_z
+\frac{\Omega}{2}\sigma_x+\frac{\delta}{2}\sigma_z,
\end{equation} 
with $\sigma_{x,z}$ being Pauli matrices.
$\Omega$ is the Rabi frequency which depends on Raman laser intensities, and $\delta$ is the detuning of the two laser beams with respect to the energy difference between the two hyperfine levels.  $H_\mathrm{int}$ denotes the contact interactions,
\begin{align}
H_\mathrm{int}[\psi]=\left(
\begin{matrix}
g|\psi_1|^2+g_{12}|\psi_2|^2 &  0\\
0 & g|\psi_2|^2+g_{12}|\psi_1|^2
\end{matrix}
\right). \notag
\end{align}
$g$ and $g_{12}$ are nonlinear coefficients for intra and intercomponent interactions respectively,
which are proportional to the corresponding $s$-wave scatting lengths.

At the first stage, we search supercurrent-carrying supersolid solutions to the GP equations. 
The wave function of supersolids can be constructed as
\begin{align}
\psi(x,t)& =\sqrt{n_0} e^{ikx-i\mu t}
\left[\begin{array}{c} \phi_1(x) \\ \phi_2(x) \end{array}\right], \notag\\
\phi_{1,2}(x)&=\sum^{L}_{j=-L} e^{i j\xi x} \phi^{(j)}_{1,2},
\label{stripe}
\end{align}
where $n_0$ denotes the mean atom density and $\mu$ is the chemical potential.  
The plane-wave prefactor $ \exp(ikx)$ is introduced for the generation of current and $k$ is the quasimomentum. $\phi_{1,2}(x)$ are periodic functions and are expressed as a superposition of plane-wave modes
 $\exp(ij \xi x)$ with coefficients $\phi^{(j)}_{1,2}$.  $L$ is the cutoff of the mode number and $\xi$ determines the period of supersolids. 
The coefficients $\phi^{(j)}_{1,2}$ satisfy the normalization condition,
\begin{align}
\sum^{L}_{j=-L}\left[ \left| \phi^{(j)}_{1}\right|^2 +\left| \phi^{(j)}_{2}\right|^2   \right] =1. \notag
\end{align}
For a given quasimomentum $k$, the constructed wave functions are fully determined by minimizing the energy functional associated to the GP equations,
\begin{align}
\mathcal{E}[\psi]&=\int dx\psi^\dagger H_{\mathrm{SOC}} \psi +\frac{g}{2}\int dx \left( |\psi_1|^4 +|\psi_2|^4  \right)    \notag\\
&\phantom{={}} +g_{12}\int dx 
|\psi_{1}|^2|\psi_{2}|^2.
\label{energy}
\end{align}
After the minimizing procedure,  the unknown quantities, i.e., $\xi$ and $\phi^{(j)}_{1,2}$, can be obtained.
The dependence of the resultant minimized energy $\mathcal{E}$ on the quasimomentum $k$, i.e., $\mathcal{E}(k)$, constitutes the dispersion relationship of supersolid stripe family.  Then substituting the resultant wave function into the GP equations, we can calculate the corresponding chemical potential,
\begin{align}
\mu&=\int dx\psi^\dagger H_{\mathrm{SOC}} \psi +g\int dx \left( |\psi_1|^4 +|\psi_2|^4  \right)    \notag\\
&\phantom{={}} +g_{12}\int dx 
|\psi_{1}|^2|\psi_{2}|^2.
\label{chemicalpotential}
\end{align}
Whether the states carry supercurrent can be checked by calculating the current density $J$,
\begin{align}
\label{current}
J&=\langle \psi|\hat{v}|\psi \rangle  \notag\\
&=n_0k+n_0\sum^{L}_{j=-L}\left[j\xi\left(|\phi^{(j)}_1|^2+|\phi^{(j)}_2|^2\right) \right.  \notag\\
&\phantom{={}}\left.+\left(|\phi^{(j)}_1|^2 - |\phi^{(j)}_2|^2 \right) \right],
\end{align}
where $\hat{v}=-i\partial_x+ \sigma_z$ is the velocity operator resulting from the the spin-orbit coupling~\cite{Mardonov2015PRL, Mardonov2015,Mardonov2018,LuoX}. 
We can see that the current density is the product of the velocity and the density.
%The current density is the product of the velocity and the density, i.e.,  $J=\langle \psi|\hat{v}|\psi \rangle $. 
The nonzero quasimomentum $k$ plays an important role in the generation of current flow.

The elementary excitation spectrum can be calculated once we know the wave function $\psi$ and associated chemical potential $\mu$.
When the system is perturbed, the total wave functions can be written as the sum of the ground-state supersolid stripe $\psi$ and perturbations $\delta \psi$, 
\begin{equation}
\Psi(x,t)=\sqrt{n_0} e^{ikx-i\mu t} \left\lbrace 
\left[\begin{array}{c} \phi_1(x) \\ \phi_2(x) \end{array}\right]+ \delta \psi (x,t) \right\rbrace ,
\end{equation}
with 
\begin{equation}
\label{perturbation}
\delta\psi(x,t)=\begin{bmatrix}
 u_{1}(x)e^{-i\omega t}
+v^{\ast}_{1}(x)e^{i\omega^* t} \\  u_{2}(x)e^{-i\omega t}
+v^{\ast}_{2}(x)e^{i\omega^* t}
\end{bmatrix}.
\end{equation}
Here,  $\omega$ is the excitation energy, and $u_{1,2}(x)$ and $v_{1,2}(x)$ are perturbation amplitudes, satisfying the normalization condition,
\begin{align}
\sum_{j=1,2}\int dx \left[|u_{j}(x)|^2-|v_{j}(x)|^2 \right]=1. \notag
\end{align}
By substituting the total wave functions into the GP equations [Eq.~(\ref{eq:motion})] and keeping the linear terms with respect to perturbation amplitudes $u_{1,2}$ and $v_{1,2}$, we obtain the Bogoliubov--de Gennes (BdG) equation~\cite{Dalfovo},
\begin{align}
\mathcal{M}\Phi=\omega\Phi,
\label{eq:BdG}
\end{align}
with
\begin{align}
\Phi=(u_{1},u_{2},v_{1},v_{2})^T,~
\mathcal{M}=\left(
\begin{array}{cc}
\mathcal{A} & \mathcal{B} \\ -\mathcal{B}^\ast & -\mathcal{A}^*
\end{array}
\right). \notag
\end{align}
The matrices $\mathcal{A}$ and $\mathcal{B}$ are
\begin{align}
\mathcal{A}&=\frac{\Omega}{2}\sigma_x+
\left(
\begin{array}{cc}
\mathcal{H}_1-i(\partial_x +ik)& n_0g_{12}\phi_1\phi^\ast_2 \\ n_0g_{12}\phi^\ast_1\phi_2 & \mathcal{H}_2+i(\partial_x +ik)
\end{array}
\right), \notag\\
\mathcal{B}&=
n_0\left(
\begin{array}{cc}
g\phi^2_1 & g_{12}\phi_1\phi_2 \\ g_{12}\phi_1\phi_2  & g\phi^2_2
\end{array}
\right), \notag
\end{align}
with
\begin{align}	
\mathcal{H}_1&=-\frac{1}{2}(\frac{\partial}{\partial x}+ik)^2+\frac{\delta}{2}+2n_0g|\phi_1|^2+n_0g_{12}|\phi_2|^2-\mu,\notag\\
\mathcal{H}_2&=-\frac{1}{2}(\frac{\partial}{\partial x}+ik)^2-\frac{\delta}{2}+2n_0g|\phi_2|^2+n_0g_{12}|\phi_1|^2-\mu.\notag
\end{align}

By diagonalizing the BdG equation, we can get the excitation energy $\omega$. Since the wave functions $\phi_{1,2}(x)$ have the period of $2\pi/\xi$, 
the matrices $\mathcal{M}$ depending on quadratic terms of $\phi_{1,2}$ are also periodic with the period $\pi/\xi$.  Therefore, the excitation spectrum will have Bloch band-gap structures and perturbation amplitudes are in the form of Bloch waves. To proceed the diagonalization of the BdG equation, we set the perturbation amplitudes as Bloch waves and expand them by the plane-wave basis,
\begin{align}
u_{1,2}(x)&=e^{iqx}\sum^{L}_{j=-L}U^{(j)}_{1,2}e^{i(2j+1)\xi x}, \notag\\
v_{1,2}(x)&=e^{iqx}\sum^{L}_{j=-L}V^{(j)}_{1,2}e^{i(2j+1)\xi x},
\label{eq:stripe-pertubation}
\end{align}
where $q$ is the perturbation quasimomentum, and $U^{(j)}_{1,2}$ and $V^{(j)}_{1,2}$ are expansion coefficients. 
The excitation spectrum $\omega(q)$ can be straightforwardly obtained by substituting Eqs.~(\ref{eq:stripe-pertubation}) into Eq.~(\ref{eq:BdG}) and diagonalizing the resultant BdG equation.  It was shown that the excitation spectrum has two gapless Nambu-Goldstone modes in the long wave-length regime ($q\rightarrow 0$) for a ground-state supersolid stripe phase and their appearance characterizes the superfluidity of the stripe phase \cite{LiY2013}.  Based on the elementary excitations, the properties such as quantum depletion~\cite{ChenX} and structure factors~\cite{ LiY2013,Chen2022} can be analysed.

The matrix $\mathcal{M}$ is non-Hermitian, allowing for the existence of imaginary excitation energies. 
In the case of imaginary energies, the perturbations in Eq.~(\ref{perturbation}) grow exponentially, so that the associated supersolid stripe is dynamically unstable.
Such dynamical instability will destroy the associated supersolid stripe.  The ground-state supersolid stripes are always dynamically stable. 
Recently, it is proposed that dynamical instability can be induced in  a metastable supersolid stripe in spin-orbit-coupled BECs~\cite{Xia}.

The energetic instability is associated with the excitation spectrum corresponding to $\tau \mathcal{M}$~\cite{Wu}, with
\begin{align}
\tau=\left(\begin{matrix} I & \\ & -I  \end{matrix}\right),\notag
\end{align}
where $I$ is a $2\times 2$ identical matrix.
The eigenvalue equation can then be written as
\begin{align}
\tau\mathcal{M}\Phi=\omega^\prime \Phi.
\label{eq:landau}
\end{align}
The matrix $\tau\mathcal{M}$ is Hermitian so that it always has a real spectrum. The energetic instability is featured with negative values of the energy spectrum.  The appearance of negative energy modes in $\omega'$ means that the associated BEC state is energetically unfavourable.  It was shown that the negative energies may happen around phonon and roton excitations for a supercurrent-carrying plane-wave state in spin-orbit-coupled BECs~\cite{Ozawa}.

%%%%%%%%%%%%%%%%%%%%%%%%%%%%%%%%%%%%%%%%%%%%%%%%%%%%%%%%%%%%%%%%%%%
\begin{figure*}[htbp]
\includegraphics[width=7in]{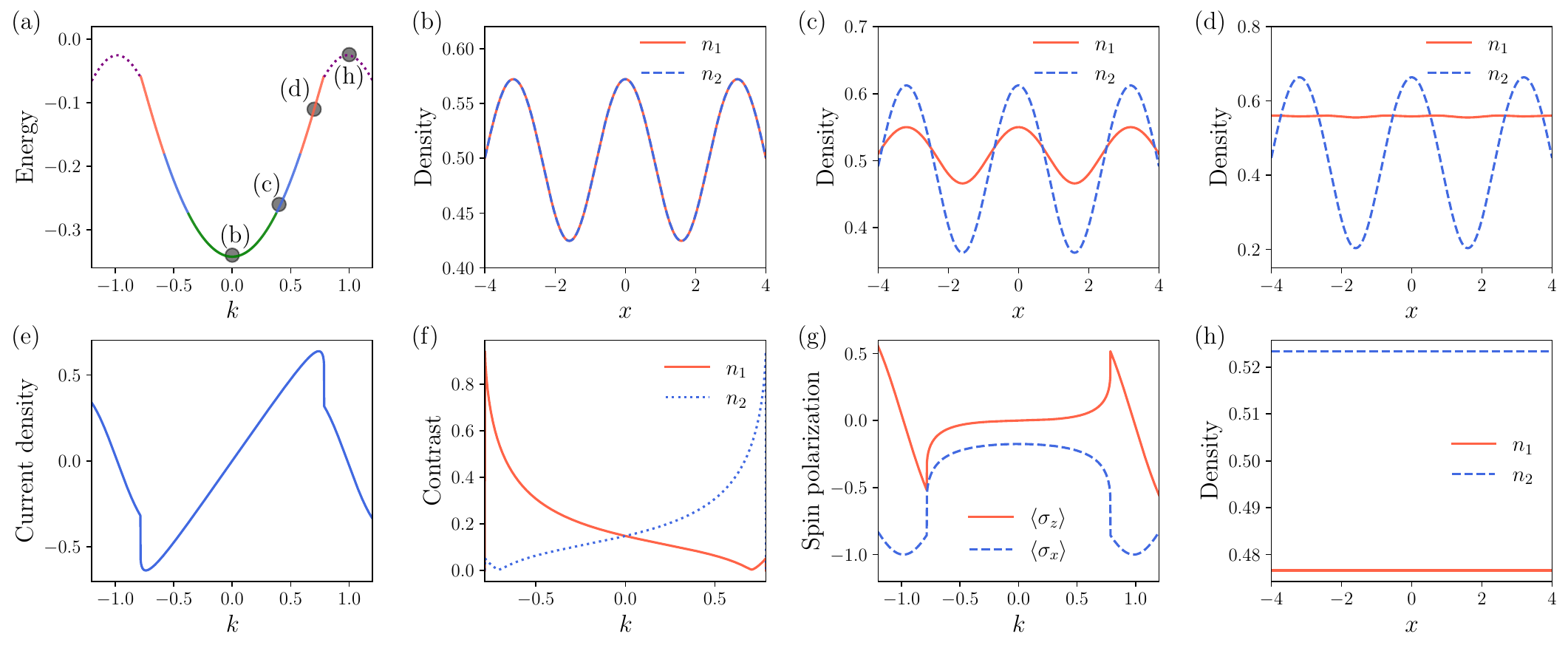}
\caption{Supercurrent-carrying supersolids in the absence of the detuning ($\delta=0$). The parameters are $\Omega=0.4$, $n_0g=0.5$, and $n_0g_{12}=0.2$.
(a) The energy of the supercurrent-carrying supersolids. In the region of $|k|<0.38$ (green line) the states are stable, in $0.38 \le |k| \le 0.58$ (blue line) they are energetically unstable, and in $0.58<|k|<0.78$ (red line) they are both dynamically and energetically unstable. Beyond $|k|=0.78$ (purple-dotted line), the states become the plane waves.  Several typical wave functions represented by labeled dots at $k=0,0.4,0.7$ are shown in (b)-(d), where the red-solid and blue-dashed lines denote the density of the first [$n_1=|\psi_1(x)|^2$] and second [$n_2=|\psi_2(x)|^2$]  components, respectively.  (e) The current density of corresponding supersolids in Eq.~(\ref{current}). (f) The contrasts of corresponding supersolids. The red-solid and blue-dashed lines are the contrasts for the first and second components, respectively. (g) The spin polarizations of corresponding supersolids. The red-solid and blue-dashed lines represent $\langle \sigma_z \rangle$ and $\langle \sigma_x \rangle$, respectively. (h) The density distribution of the plane wave at $k=1$.}
\label{fig1}
\end{figure*}
%%%%%%%%%%%%%%%%%%%%%%%%%%%%%%%%%%%%%%%%%%%%%%%%%%%%%%%%%%%%%%%%%%%

%%%%%%%%%%%%%%%%%%%%%%%%%%%%%%%%%%%%%%%%%%%%%%%%%%%%%%%%%%%%%%%%%%%
\begin{figure*}[t]
\includegraphics[width=6.5in]{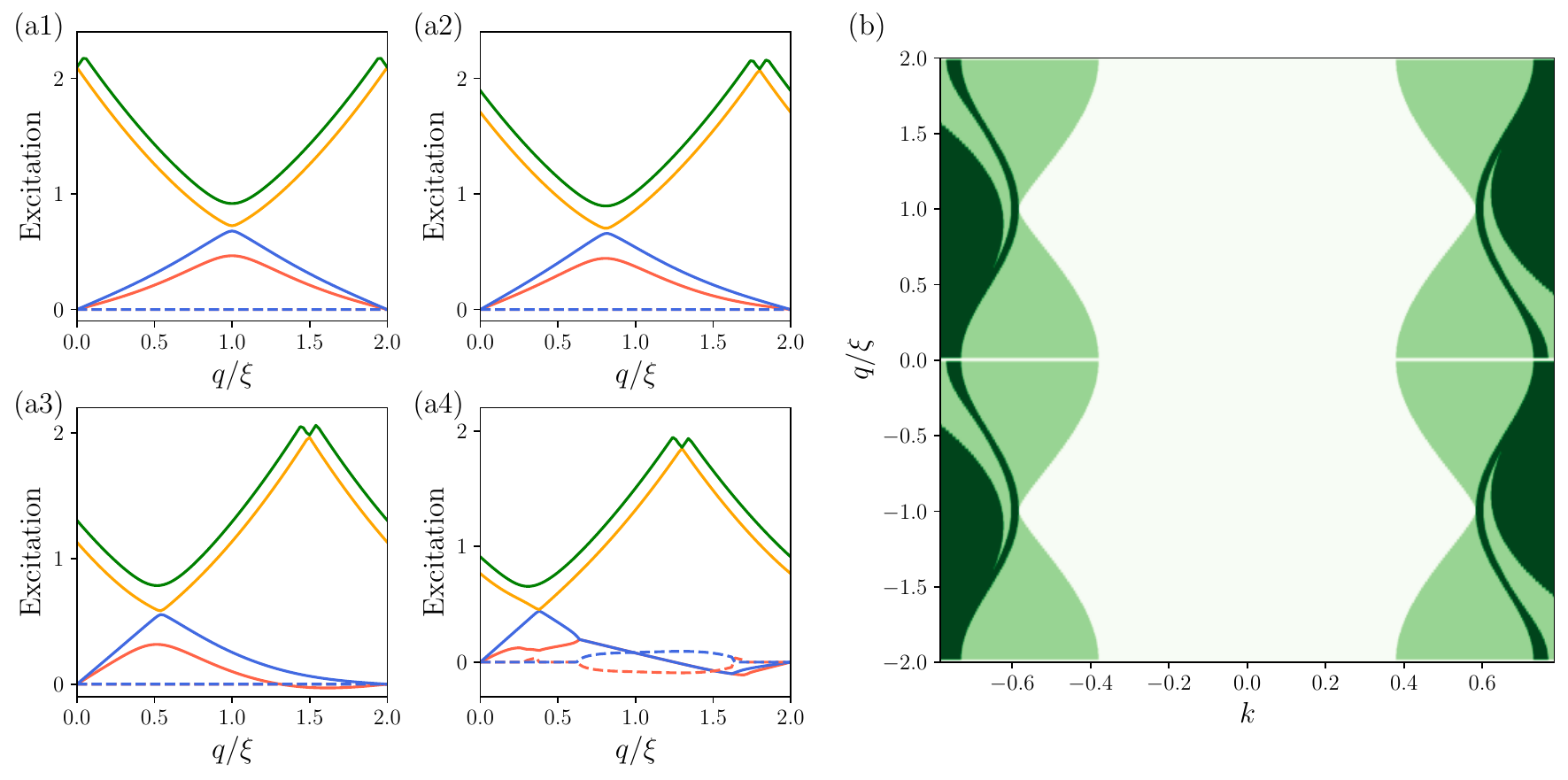}
\caption{Elementary excitation spectrum $\omega(q)$ and instabilities of supercurrent-carrying supersolids in the absence of the detuning ($\delta=0$). The parameters are $\Omega=0.4$, $n_0g=0.5$, and $n_0g_{12}=0.2$.
[(a1)-(a4)] Elementary excitation spectrum of the associated supersolids at $k=0,0.2,0.5,0.7$. Only the four lowest bands are shown. 
The solid and dashed lines are the real and imaginary parts of the spectra, respectively.
(b) The instability regions in the ($k,q$) plane. The white, light-green, and dark-green regions represent stable, energetically unstable and dynamically unstable modes, respectively.}
\label{fig2}
\end{figure*}
%%%%%%%%%%%%%%%%%%%%%%%%%%%%%%%%%%%%%%%%%%%%%%%%%%%%%%%%%%%%%%%%%%%

\section{Supercurrent-carrying supersolid stripes without the detuning}
\label{nodetuning}

In spin-orbit-coupled BEC experiments, the detuning $\delta$ is an important tunable parameter. With or without the detuning, the symmetries of $H_{\mathrm{SOC}}$ changes dramatically, which will lead to a profound impact on relevant physics. We first study the existence and stability of supercurrent-carrying supersolids in the absence of the detuning.   It is already known that the ground-state supersolid stripe phase which does not carry supercurrent only exists in the miscible interactions $g>g_{12}$~\cite{LiY,Martone2014,LiJR}. This motivates us to study the supercurrent-carrying analogs in this interaction regime with an intuitive expectation that they may share the same stability with the ground states if they exist.  
We choose $n_0g=0.5$, $n_0g_{12}=0.2$, and $\Omega=0.4$ in all the numerical calculations, and thus the supersolid stripe at $k=0$ is the ground state of the system.

By applying the trial wave functions in Eq.~(\ref{stripe}) and the minimization procedure illustrated in the previous section, we get the supercurrent-carrying supersolid states, 
whose dispersion relation $\mathcal{E}(k)$ is shown in Fig.~\ref{fig1}(a). 
The dispersion of the stripe family has a parabolic-like shape. The supersolid stripe with $k=0$ has the lowest energy, corresponding to the ground state of the system, which is widely studied in literature.  Its density crystalline structures are demonstrated in Fig.~\ref{fig1}(b). The two components have the same periodic structure, i.e., $|\psi_1(x)|^2=|\psi_2(x)|^2$. Therefore, its spin polarization is $\langle \sigma_z \rangle=\int dx ( |\psi_1|^2-|\psi_2|^2)=0$. 
The period of crystalline densities is $\pi/\xi$ with $\xi=0.98$. Such value of $\xi$ is very close to the spin-orbit-coupling strength which is $1$ in our dimensionless units.
The supersolid family demonstrated by the solid line in Fig.~\ref{fig1}(a) has the same period with the $k=0$ supersolid. Two typical density profiles at $k=0.4$ and $0.7$ are depicted in Figs.~\ref{fig1}(c) and \ref{fig1}(d), respectively. Compared with the $k=0$ supersolid, the family has three distinguished features.

(1)  The supersolid family does carry current flow.  The calculated current density from Eq.~(\ref{current}) for the associated supersolids with different $k$ is shown in  Fig.~\ref{fig1}(e).  Only the $k=0$ supersolid has a zero current density, and all others are nonzero.  The sign of the current density is same as the quasimomentum $k$. For $k>0$ states, the current flows in the positive direction. In contrast, for $k<0$ states, the current density is negative, which means that it flows towards the negative direction.  The nonzero current witnesses that it is essentially important to introduce the plane-wave prefactor $\exp(ikx)$ in the construction of supersolid wave functions. 

(2) The contrasts of crystalline densities are spin-dependent and vary obviously depending on the quasimomentum $k$. 
The contrast is defined as
\begin{align}
C=\frac{n_{\mathrm{max}}-n_{\mathrm{min}}}{n_{\mathrm{max}}+n_{\mathrm{min}}},
\end{align}
where $n_{\mathrm{max}}$ ($n_{\mathrm{min}}$) is the maximum (minimum) of the density. From the density profiles in Figs.~\ref{fig1}(c) and \ref{fig1}(d), we can see that two components still overlap with each other, while they have apparently different contrasts. The contrasts of the two components as a function of the quasimomentum $k$ are shown in Fig.~\ref{fig1}(f).  For $k>0$, the contrast of the first component decreases (the red-solid line), and that of the second component increases (the blue-dashed line) 
by raising $k$.
However, the dependence of $C$ on $k$ changes oppositely  for $k<0$.  When $|k|$ reaches a critical value, the contrast of the one of two components completely disappears, and this component turns into a plane wave.  A bit beyond the critical $|k|$, the supersolid family can not exist and the solutions become the plane-wave phase which is represented by the purple-dotted line in Fig.~\ref{fig1}(a).  A typical plane-wave density distribution is plotted in Fig.~\ref{fig1}(h).  The obvious feature of the plane-wave phase is its spin polarization, which is a result of the spontaneous breakdown of  time-reversal symmetry. 

(3) Most of the supercurrent-carrying supersolid states are spin-balanced.  At first glance on Figs.~\ref{fig1}(c) and \ref{fig1}(d), two components may be spin-imbalanced. However, the density oscillating centers of the two components are not  coincident, resulting in the spin balance between the two components. 
The calculated spin polarization $\langle \sigma_z \rangle$ and $\langle \sigma_x \rangle$ are demonstrated in Fig.~\ref{fig1}(g).  
It is shown that the two components are spin-balanced ($\langle \sigma_z \rangle\approx0$) in the dominated region of the quasimomenta ($|k|<0.7$).
When $|k|>0.7$, the supersolids change dramatically to adjust distributions including contrast and spin polarization for the preparation of transition to the plane-wave phase.

The three features are intrinsically correlated, which can be understood in the following way. We start from the GP equation and neglect the nonlinear interactions, $H_{\mathrm{SOC}} |\psi\rangle=\mathcal{E} |\psi\rangle$, with $|\psi\rangle$ being the supersolid stripe state and the associated energy $\mathcal{E}$.  The stripe wave function is Bloch-wave-like, i.e., $|\psi\rangle= e^{ikx} |\phi\rangle$ with  $|\phi\rangle$ being a periodic function.  Then the GP equation becomes $ H' |\phi\rangle=\mathcal{E} |\phi\rangle  $. Here,   $H'=e^{-ikx} H_{\mathrm{SOC}} e^{ikx}= -(\partial_x +ik)^2/2-i(\partial_x+ik)\sigma_z +\Omega \sigma_x/2$.   By applying the Hellman-Feynman theorem, we get $\partial \mathcal{E} /\partial k= \langle \phi |\partial H'/\partial k | \phi\rangle = k +\langle \phi |(-i\partial_x) | \phi\rangle+\langle \phi |\sigma_z | \phi\rangle = J$~\cite{Pethick}. Meanwhile, the spin polarization of the supersolids is zero, $\langle \phi |\sigma_z | \phi\rangle =0$, as shown in Fig.~\ref{fig1}(g), and $ \langle \phi |(-i\partial_x) | \phi\rangle \approx 0$. We then have  $J\approx k$ and  $\partial \mathcal{E} /\partial k\approx k$.
 Therefore, the current density linearly depends on the quasimomentum $k$ and its sign is relevant to the sign of  the latter. This provides an insight for the understanding of the results in Fig.~\ref{fig1}(e).   $\partial \mathcal{E} /\partial k\approx k$ leads to $\mathcal{E} \propto k^2$ which explains the parabolic-like shape of dispersion relation shown in Fig.~\ref{fig1}(a).  
On the other hand, when $k=0$, the Hamiltonian $H'$ has the time-reversal symmetry, i.e., $\mathcal{T} H'\mathcal{T}^{-1}=H'$, 
with $\mathcal{T}=\mathcal{K}\sigma_x$ and $\mathcal{K}$ being the complex conjugate operator. 
The $k=0$ supersolid inherits the time-reversal symmetry, giving rise to $|\psi_1(x)|^2=|\psi_2(x)|^2$.
Therefore the two components have the same contrast. However, the presence of nonzero quasimomentum breaks the time-reversal symmetry.  Consequently, the two components in the supercurrent-carrying supersolids have different contrasts.

Given that the supercurrent-carrying supersolid stripes can exist, we study their elementary excitations by solving the BdG equations [Eq.~(\ref{eq:BdG})] with the purpose for examining their dynamical instabilities, and we also calculate  Eq.~(\ref{eq:landau}) to examine energetic instabilities. The elementary excitations $\omega(q)$  
for the supersolid states with $k=0,0.2,0.5,0.7$  are shown in Figs.~\ref{fig2}(a1)-\ref{fig2}(a4).  As the supersolids are periodic in densities with the period of $\pi/\xi$, the elementary excitation spectra have Bloch band-gap structures with the Brillouin zone size $2\xi$.  For the supersolid with $k=0$, the excitation spectrum is symmetric with respect to the Brillouin zone centered at $q=0$ [Fig.~\ref{fig2}(a)].
The typical feature is that the lowest two bands have two gapless Nambu-Goldstone modes in the long wave-length region ($q\rightarrow 0$).  
The time-reversal symmetry is broken by the nonzero quasimomentum $k$. 
Consequently, the excitation spectrum losses the symmetry with respect to the Brillouin center.  The excitation spectrum of the $k=0.2$ supersolid in Fig.~\ref{fig2}(a2) shows such an asymmetry. The energy of two gapless modes softens in the $q<0$ region (i.e., around $q/\xi=2$ in the figure). There is no imaginary-energy excitation in both $k=0$ and $k=0.2$ supersolids, which means that these two supersolids are dynamically stable.   
For the supersolid with $k=0.5$, the softening of the gapless modes around $q/\xi=2$ leads to the lowest mode becoming energy-negative [see Fig.~\ref{fig2}(a3)]. It was proved that the energetic instability can also be indicated from the existence of negative-energy modes in elementary excitation based on the BdG equation instead of calculating Eq.~(\ref{eq:landau})~\cite{ChenZ2010}. The appearance of negative-energy modes shown in Fig.~\ref{fig2}(a3) is a signal that the corresponding $k=0.5$ supersolid is energetically unstable. However, this supersolid is dynamically stable since there is no imaginary energy excitation.  The excitation spectrum demonstrated in Fig.~\ref{fig2}(a4) for the $k=0.7$ supersolid is very different from others in Figs.~\ref{fig2}(a1)-\ref{fig2}(a3). There are imaginary-energy excitations (represented by dashed lines in the figure), which indicates the $k=0.7$ supersolid dynamically unstable.  

The imaginary-energy and negative-energy excitations emerge in the lowest two bands of the elementary excitation spectra, and they are confined in a finite quasimomentum $q$ region. We systematically identify the region of the quasimomentum $q$ where imaginary-energy and negative-energy excitations happen by calculating Eqs.~(\ref{eq:BdG}) and (\ref{eq:landau}), respectively. The results are shown in the ($k,q$) plane in Fig.~\ref{fig2}(b).
In this figure, the white region represents that there is no imaginary-energy and negative-energy excitation, there are negative-energy excitations in the light-green region, and in the dark-green region there are imaginary-energy excitations.  
The structures in Fig.~\ref{fig2}(b) have an exact symmetry $(k,q)\rightarrow (-k,-q)$.  This is because the BdG equations in Eq.~(\ref{eq:landau}) possess the symmetry $\mathcal{M}(k,q)=\mathcal{M}(-k,-q)$.  Appearance of imaginary-energy (negative-energy) excitations means that the associated supersolids are dynamically  (energetically) unstable. From Fig.~\ref{fig2}(b), we can demarcate instability regions: the supersolids in $|k|<0.38$ are stable, in $0.38 \le |k| <  0.58$ are energetically unstable, and in $0.58 \le |k| < 0.78$ are dynamically and energetically unstable. We also incorporate the instability regions by different colored lines in Fig.~\ref{fig1}(a). 

%%%%%%%%%%%%%%%%%%%%%%%%%%%%%%%%%%%%%%%%%%%%%%%%%%%%%%%%%%%%%%%%%%%
\begin{figure}[t]
\includegraphics[width=3.4in]{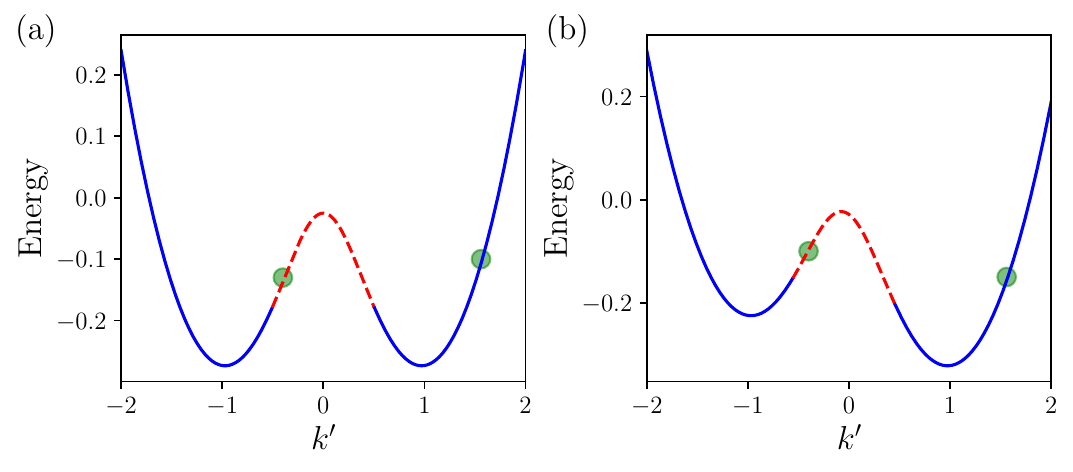}
\caption{(a) Energy of the plane-wave phase in the absence of detuning ($\delta=0$). The parameters are the same as Figs.~\ref{fig1} and~\ref{fig2}. 
The red-dashed line in the region of $|k'|<0.51$ indicates that the corresponding plane waves are dynamically unstable. Dynamically stable plane waves are represented by the blue solid line.   (b) Energy of the plane-wave phase in the presence of the detuning ($\delta=0.1$). The parameters are the same as Figs.~\ref{fig4} 
and~\ref{fig5}. The plane waves in the region $-0.56<k'<0.46$ (red-dashed line) are dynamically unstable.  Superposition of two 
plane waves (labeled by green dots) may give a main contribution to the corresponding supersolid.	}
\label{fig3}
\end{figure}
%%%%%%%%%%%%%%%%%%%%%%%%%%%%%%%%%%%%%%%%%%%%%%%%%%%%%%%%%%%%%%%%%%%

The above identified energetic instability can be understood straightforwardly. The velocity of carried supercurrent in supersolids is proportional to the current density which is approximately $J\approx n_0 k$. As the quasimomentum $|k|$ increasing from zero, the supercurrent velocity also increases. 
When $|k|$ reaches a critical value, the supercurrent-carrying ability is broken according to Landau's criterion of superfluidity, which leads to energetic instability. Therefore, the supercurrent-carrying supersolids are energetically stable only within a certain quasimomentum $k$ region around $k=0$.

%%%%%%%%%%%%%%%%%%%%%%%%%%%%%%%%%%%%%%%%%%%%%%%%%%%%%%%%%%%%%%%%%%%
\begin{figure*}[t]
\includegraphics[width=6.5in]{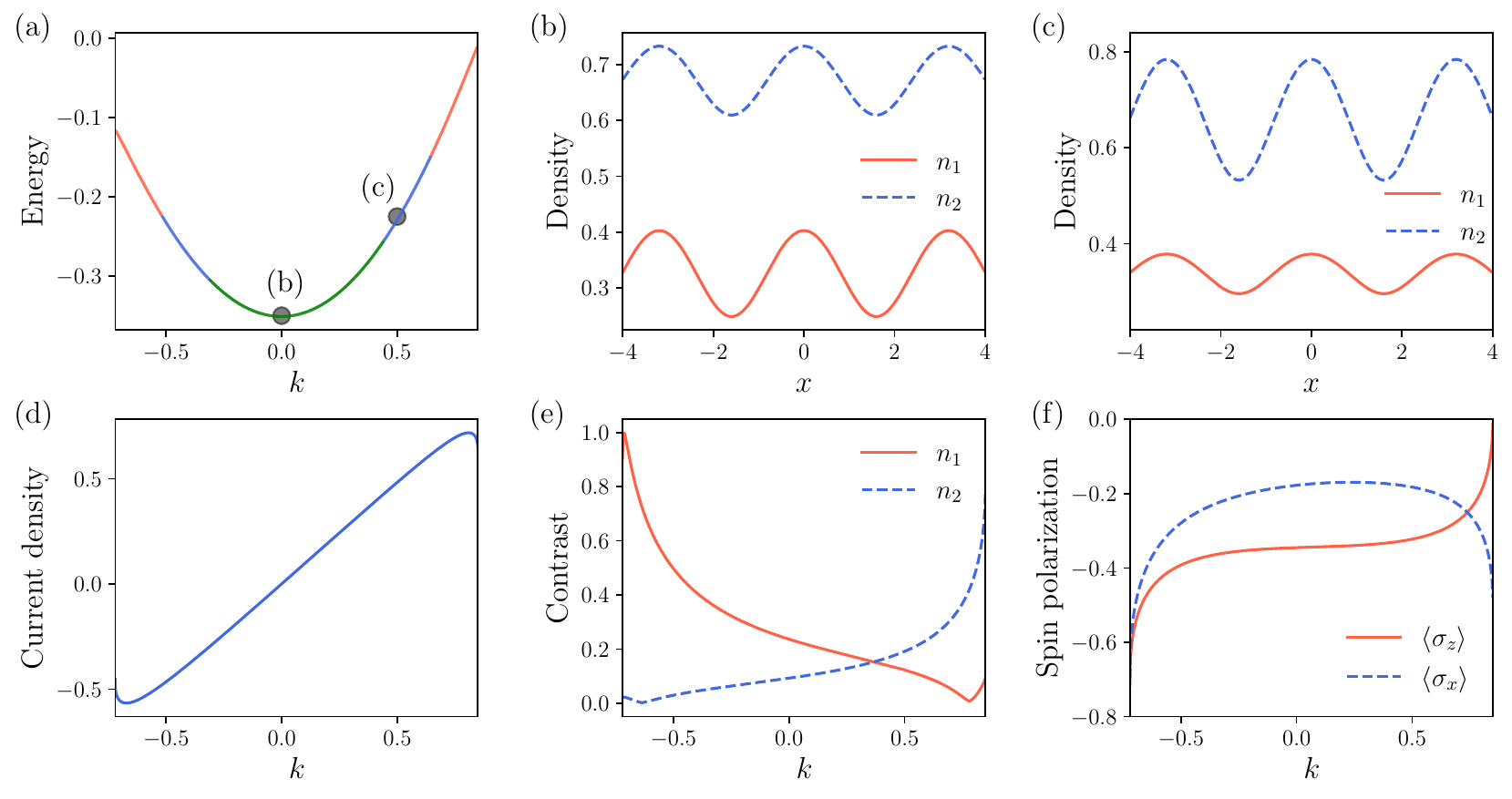}
\caption{Supercurrent-carrying supersolids in the presence of the detuning ($\delta=0.1$).  The parameters are $\Omega=0.4$, $n_0g=0.5$, and $n_0g_{12}=0.2$.
(a) The energy of the supercurrent-carrying supersolids.  In the region of $-0.33<k<0.35$ (green line) the states are stable, in $-0.52 \le k \le -0.33$ and $0.35 \le k \le 0.58$ (blue line) they are energetically unstable, and in $k>0.58$ and $k<-0.52$ (red line) they are both dynamically and energetically unstable. Typical wave functions labeled by dots at $k=0$ and 0.5 are shown in (b) and (c), where the red-solid and blue-dashed lines denote the density of the first [$n_1=|\psi_1(x)|^2$] and second [$n_2=|\psi_2(x)|^2$]  components respectively.  (d) The current density of corresponding supersolids in Eq.~(\ref{current}). (e) The contrasts of corresponding supersolids. The red-solid and blue-dashed lines are the contrasts of the first and second components, respectively. (f) The spin polarizations of corresponding supersolids. The red-solid and blue-dashed lines represent $\langle \sigma_z \rangle$ and $\langle \sigma_x \rangle$, respectively.}
\label{fig4}
\end{figure*}
%%%%%%%%%%%%%%%%%%%%%%%%%%%%%%%%%%%%%%%%%%%%%%%%%%%%%%%%%%%%%%%%%%%

The dynamical instability of the supercurrent-carrying supersolids identified above is of interest. We provide an insightful picture to qualitatively understand  its emergence. 
It is noticed that the supersolids are composed of plane waves $\exp[i(k+j\xi)
x]$ with $j$ being an integer [see Eq.~(\ref{stripe})]. This motivates us to study the pure plane-wave solutions firstly. The wave function of plane-wave phase is~\cite{Ozawa}
\begin{equation}
\psi=\sqrt{n_0} e^{ik'x-i\mu' t}\begin{pmatrix}
\varphi_1 \\ \varphi_2
\end{pmatrix}.
\end{equation}
Here the spinor $(\varphi_1,\varphi_2)^T$ is spatially independent and satisfies $|\varphi_1|^2+|\varphi_2|^2=1$.  The spinor together with the chemical potential $\mu'$ can be determined by minimizing the associated energy functional. 
The resultant energy $\mathcal{E}'(k')$ of the plane-wave phase is shown in Fig.~\ref{fig3}(a), 
which has a similar structure to the single-particle spin-orbit-coupled energy. The analysis of dynamical instability of the plane-wave phase can be performed by solving the associated BdG equations similar to Eq.~(\ref{eq:BdG}). The result is that the plane waves in the region of $|k'|<0.51$ are dynamically unstable [represented by the red-dashed line in Fig.~\ref{fig3}(a)], and others are dynamically stable. Such dynamical instability of the plane-wave phase was revealed to relate with the negative effective mass~\cite{Ozawa}. We interpret the dynamical instability of the supercurrent-carrying supersolids from the instability of the plane-wave phase. 
In the wave function of  the supersolids $\sum_j \exp[i(k+j\xi)
x] (\phi_1^{(j)}, \phi_2^{(j)})^T$, the $j=\pm 1$ plane waves have a dominant occupation, which can be seen from the numerically calculated wave function.  When $k\ge 0.47$, the $j=-1$ plane wave $\exp[i(k-\xi)
x]$ enters into the dynamically unstable region of the pure plane-wave phase (since $\xi=0.98$) [see the green dots in Fig.~\ref{fig3}(a)], the consequence of which is that the whole wave function of the supersolids may become dynamically unstable.  In the same way, when $k\le -0.47$, the $j=1$ plane wave  $\exp[i(k+\xi)
x]$ accesses to the dynamically unstable region of the pure plane-wave phase. Therefore, the supersolids are expected to be dynamically unstable when $|k|\ge0.47$.  This predicted region is different from the previous numerical result ($|k|\ge0.58$). This indicates that the interpretation of the dynamical instability of the supersolids by that of the pure plane-wave phase is not exact. After all, the supersolids are an assembly of many plane waves. Nevertheless, it provides a profound insight to understand the dynamical instability of the supercurrent-carrying supersolids.

Finally, we emphasize that it is the plane-wave factor $\exp(ikx)$ that induces supercurrent and instability in supersolid stripes. The stripe at $k=0$ is the ground state of the system, which does not carry current and is always dynamically and energetically stable.  The energetic instability indicates that the associated supersolid stripe is energetically unfavourable, and the system actually prefers the plane-wave phase~\cite{Zhang}. The dynamical instability breaks the associated supersolid stripe to give rise to complicate and random density patterns. The fate of such an unstable stripe is similar to that of a dynamically unstable Bloch wave in optical lattices~\cite{Wu2003,Konotop}, where the loss of atoms and random density patterns have been experimentally observed~\cite{Fallani,DeSarlo}. Meanwhile, the dynamical instability of the Bloch wave is used as a signal for the existence of solitonic solutions~\cite{Konotop2004,Kartashov}.

%%%%%%%%%%%%%%%%%%%%%%%%%%%%%%%%%%%%%%%%%%%%%%%%%%%%%%%%%%%%%%%%%%%
\begin{figure*}[t]
\includegraphics[width=6.5in]{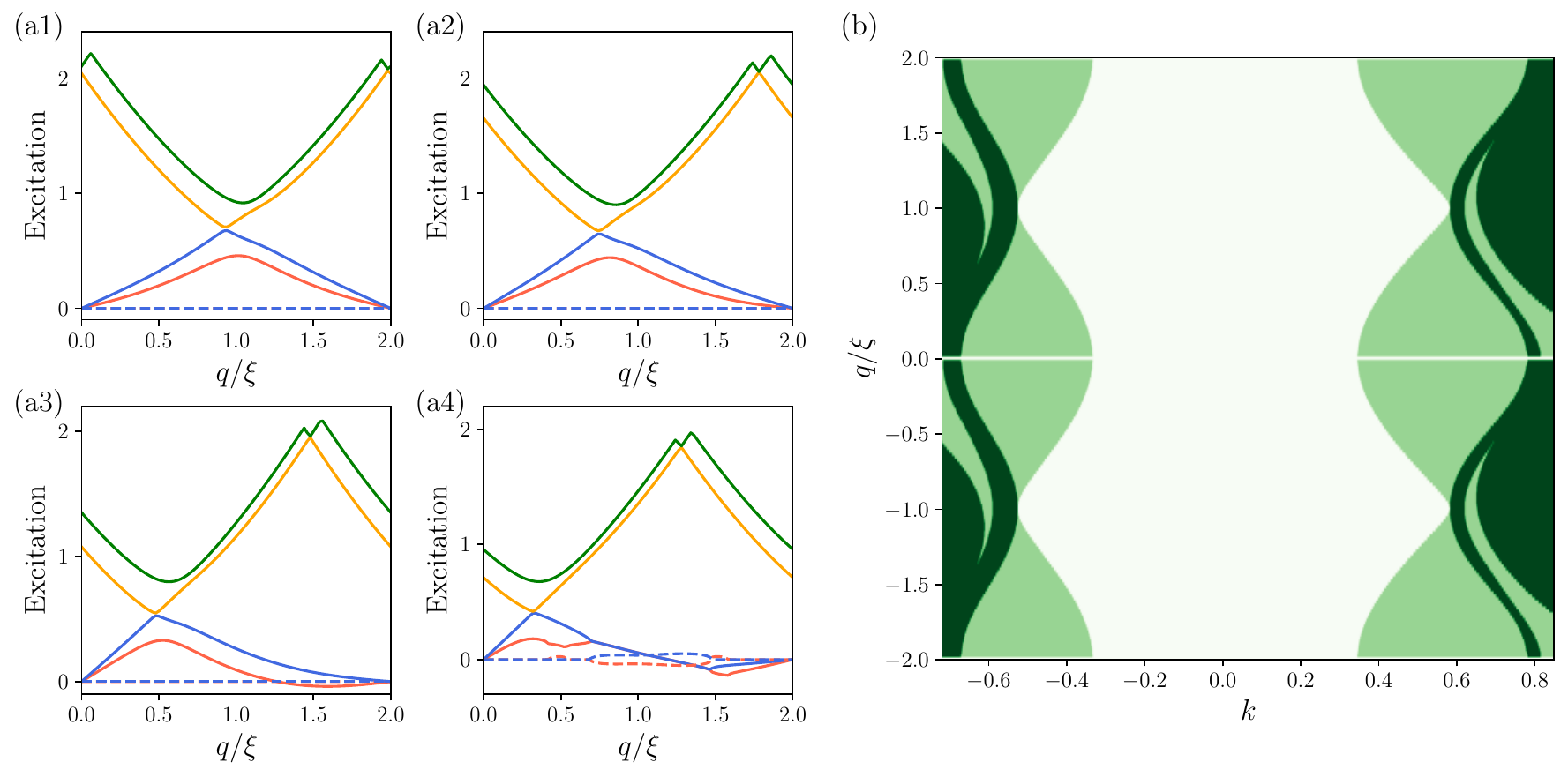}
\caption{Elementary excitation spectrum $\omega(q)$ and instabilities of supercurrent-carrying supersolids in the presence of the detuning ($\delta=0.1$). The parameters are $\Omega=0.4$, $n_0g=0.5$, and $n_0g_{12}=0.2$.
[(a1)-(a4)] Elementary excitation spectrum of the associated supersolids at $k=0,0.2,0.5,0.7$. Only the four lowest bands are demonstrated. The solid and dashed lines are the real and imaginary parts of the spectra, respectively.
(b) The instability regions in the ($k,q$) plane. The white, light-green, and dark-green regions represent stable, energetically unstable and dynamically unstable modes, respectively.}
\label{fig5}
\end{figure*}
%%%%%%%%%%%%%%%%%%%%%%%%%%%%%%%%%%%%%%%%%%%%%%%%%%%%%%%%%%%%%%%%%%%

\section{Supercurrent-carrying supersolid stripe with the detuning}
\label{detuning}

In the presence of the detuning term $(\delta/2) \sigma_z$,  $H_{\mathrm{SOC}}$ does not have the time-reversal symmetry,
and the system prefers spin-polarized solutions with nonzero $\langle \sigma_z \rangle$. 
The experimentally tunable detuning brings novel physics in spin-orbit-coupled BECs~\cite{Zheng,Hamner2014}.
We will show that supercurrent-carrying supersolid stripes can still exist in the presence of the detuning.  Using the same constructed wave functions in Eq.~(\ref{stripe}) and following the procedure of minimizing the associated energy functional, we obtain the supersolid stripes in the parameter regime of $\delta=0.1$, $\Omega=0.4$, $n_0g=0.5$, and $n_0g_{12}=0.2$. 

The dispersion relation $\mathcal{E}(k)$ is demonstrated in Fig.~\ref{fig4}(a). The whole family has the same density period $\pi/\xi$ with $\xi\approx0.98$, which is nearly the same as the zero-detuning case in the previous section. The density distributions for the $k=0$ and $k=0.5$ supersolids are described in Figs.~\ref{fig4}(b) and~\ref{fig4}(c), respectively.  The outstanding feature is that the two components are spin-imbalanced.  The second component has a larger occupation when $\delta>0$.  The family carries supercurrent and the current density is shown in Fig.~\ref{fig4}(d). It is obvious that the current density is still linearly dependent on the quasimomentum, $J\propto k$. The contrast of each component as a function of the quasimomentum is demonstrated in Fig.~\ref{fig4}(e).  In the positive quasimomentum region, the contrast of the first component decreases with $k$ increasing, 
and it completely disappears after $k$ reaching a critical value, indicating that the first component turns to be a plane wave. Beyond the critical quasimomentum, the supersolids can not exist. Similarly, in the negative quasimomentum region, the contrast of the second component gradually disappears. Therefore, the supersolids can only exist in a finite quasimomentum region, which is $-0.72\leqslant k\leqslant 0.85$.  Such an asymmetric existence region with respect to $k=0$ becomes a main difference compared with the zero-detuning case shown in Fig.~\ref{fig1}(a).  In the main part of the existence region, the spin polarizations $\langle \sigma_z\rangle$ and $\langle \sigma_x\rangle$  are almost constant and negative [see Fig.~\ref{fig4}(f)]. Around the existence region boundaries, the spin polarizations change dramatically.

The breakdown of supercurrent-carrying ability may be induced by energetic and dynamical instabilities. In order to analyze the instability, we calculate elementary excitations of supercurrent-carrying supersolid stripes.  The elementary excitation spectrum $\omega(q)$ for the supersolids with $k=0,0.2,0.5,0.7$ are shown in Figs.~\ref{fig5}(a1)-(a4), respectively. All the spectra show the asymmetry with respect to $q=0$ due to the lack of the time-reversal symmetry, which is a result of  the nonzero detuning. The structures of the spectra are similar to these in the zero-detuning cases shown in  Figs.~\ref{fig2}(a1)-\ref{fig2}(a4).
The negative-energy excitations [Fig.~\ref{fig5}(a3)] and the imaginary-energy excitations  [Fig.~\ref{fig5}(a4)] happen in the lowest two bands. From the elementary excitation spectrum, we capture instability regions and show them in the $(k,q)$ plane in Fig.~\ref{fig5}(b).  The instability regions (shadowed areas in the figure) lose the symmetry  of $(k,q)\rightarrow (-k,-q)$ since in the BdG equations $\mathcal{M}(k,q)\ne \mathcal{M}(-k,-q)$ in presence of the detuning.  From boundaries of light-green regions (for energetic instability) and dark-green regions (for dynamical instability), we conclude that in the region $-0.33<k<0.35$ [shown by the green line in Fig.~\ref{fig1}(a)] the supersolids are stable,  in $-0.52 \le k \le -0.33$ and $0.35 \le k \le 0.58$ [the blue line in Fig.~\ref{fig1}(a)] they are energetically unstable, and in $k>0.58$ and $k<-0.52$ [the red line in Fig.~\ref{fig1}(a)] they are both dynamically and energetically unstable.

The onsets of the dynamical instability are at $k=0.58$ and $k=-0.52$, which are not symmetric with respect to $k=0$. 
The occurring of the dynamical instability as well as its asymmetric parameter region can be explained from the pure plane-wave phase.   The energy of the plane-wave phase shown in Fig.~\ref{fig3}(b) still has a double-well structure. However, in comparison with the zero-detuning case shown in Fig.~\ref{fig3}(a), the nonzero detuning biases the double wells, so that the dispersion loses the symmetry with respect to $k'=0$. Furthermore, the region of the dynamical instability in the plane-wave phase also loses the symmetry with respect to $k'=0$, and it is $-0.56<k'<0.46$ [represented by the red-dashed line in Fig.~\ref{fig3}(b)]. Even in the presence of the detuning, the $j=\pm 1$ plane waves in the wave function of supercurrent-carrying supersolids $\sum_j \exp[i(k+j\xi)
x] (\phi_1^{(j)}, \phi_2^{(j)})^T$ still have a dominant occupation. Considering $\xi\approx0.98$ in the supersolid family, when $k>0.42$, the $j=-1$ plane wave $ \exp[i(k-\xi)
x] $ enters into the dynamically unstable region of the pure plane-wave phase [see the green dots in Fig.~\ref{fig3}(b)]. 
Similarly, when $k<-0.52$, the $j=1$ plane wave  $\exp[i(k+\xi)x]$ enters into the dynamically unstable region of the pure plane-wave phase. Therefore, intuitively following the dynamical instability of the pure plane-wave phase, we predict the supersolid family becomes dynamically unstable in the region of $k>0.42$ and $k<-0.52$. The predicted onsets are at $k=0.42$ and $k=-0.52$, which are not symmetric with respect to $k=0$. The reason of such asymmetry is that the dispersion relation of the pure plane-wave phase does not have the symmetry with respect to $k'=0$ as shown in Fig.~\ref{fig3}(b) due to the detuning.  It is noticed that the predicted onsets are not exact to the numerically calculated values.

\section{Experimental considerations}
\label{exp}

We have uncovered the existence of supercurrent-carrying supersolid stripes. Now we discuss their experimental accessibility. Experiments may start from a two-component BEC with miscible interactions in an elongated trap. Dressing the two-component BEC along the longitudinal direction by adiabatically ramping up two Raman lasers can prepare a supersolid stripe~\cite{LiJR}.  
Such an implementation of supersolid stripes corresponds to the ground state, i.e., the $k=0$ supersolid, which does not carry supercurrent. 
Then the $k=0$ supersolid may be accelerated by applying an external force $F$ along the longitudinal direction, such as the gravity-induced force~\cite{Olson2014,Olson2017}. The quasimomentum of the supersolid will increase linearly with the time $t$, i.e., $k=Ft/\hbar$~\cite{Beek}. By controlling the duration of the applied force, we can prepare the supersolid with a desired quasimomentum. 
Once the supercurrent-carrying supersolids are prepared, their dynamical instability can be observed by holding the supersolids for a certain time and measuring the loss of condensed atoms~\cite{Hamner,Fallani}.
In the spin-orbit-coupled supersolid stripe experiment in Ref.~\cite{LiJR}, the excitations in the transverse directions are completely irrelevant to the observed results. 
We expect that the transverse degrees of freedom may not qualitatively change the predicted instability results of the supercurrent-carrying supersolids,
which are reminiscent of Bloch states in a longitudinal optical lattice.
It has been shown that the transverse degrees of freedom do not qualitatively modify the instabilities of the Bloch states existing along the longitudinal direction~\cite{Modugno}.

\section{Conclusions}
\label{conclusion}

By constructing Bloch-wave-like supersolid wave functions, we revealed the existence of supercurrent-carrying supersolid stripes in spin-orbit-coupled BECs. The family of supersolid stripes has the same density period and possesses a parabolic-like dispersion. The current carried by these supersolids is proportional to the quasimomentum. 
Energetic and dynamical instabilities emerge with a large current, which limit the supercurrent-carrying ability of the supersolids
and even destroy them.
The energetic instability relates to the well-known Landau's criterion of superfluidity.  The dynamical instability is interpreted as that the supersolid stripes involve a dynamically unstable plane wave.
These results provide a possible route to exploring exotic supersolids carrying supercurrents and their instabilities.

\section*{Acknowledgment}
We are grateful to Prof.~Tomoki Ozawa for providing us the insight into the explanation of the dynamical instability of the supercurrent-carrying supersolids from that of the pure plane-wave phase.  We also acknowledge valuable discussions with Prof.~Xi-Wang Luo.
This work is supported by National Natural Science Foundation of China with Grants No.~12374247 and No.~11974235 
and Shanghai Municipal Science and Technology Major Project (Grant No. 2019SHZDZX01-ZX04).
L.H. also acknowledges support from Okinawa Institute of Science and Technology Graduate University. Q.Z. is supported by NKRDPC (Grant No. 2022YFA1405304), NSFC (Grant No. 12004118), and Guangdong Basic and
Applied Basic Research Foundation (Grants No. 2020A1515110228 and No.
2021A1515010212).

\bibliography{SC}

\end{document}